\documentclass[nofootinbib,superscriptaddress,a4paper, twocolumn, floatfix]{revtex4-2}
\usepackage[english]{babel}
\usepackage[utf8]{inputenc}
\usepackage[colorinlistoftodos, color=green!40, prependcaption]{todonotes}
\usepackage{amsthm}
\usepackage{mathtools}
\usepackage{physics}
\usepackage{braket}
\usepackage{listings}
\usepackage{xcolor}
\usepackage{graphicx}
\usepackage[left=23mm,right=13mm,top=35mm,columnsep=15pt]{geometry} 
\usepackage{adjustbox}
\usepackage{placeins}
\usepackage[T1]{fontenc}
\usepackage{lipsum}
\usepackage{csquotes}
\usepackage[pdftex, pdftitle={Article}, pdfauthor={Author}]{hyperref} 
\usepackage{booktabs}
\usepackage{amssymb}
\usepackage{bm}
\usepackage[ruled,vlined]{algorithm2e}
\usepackage[caption=false]{subfig}

\begin{document}

\title{Quantum compression with classically simulatable circuits}

\author{Abhinav Anand}
\email[E-mail:]{abhinav.anand@mail.utoronto.ca}
\affiliation{Chemical Physics Theory Group, Department of Chemistry, University of Toronto, Canada.}

\author{Jakob~S.~Kottmann}
\affiliation{Chemical Physics Theory Group, Department of Chemistry, University of Toronto, Canada.}
\affiliation{Department of Computer Science, University of Toronto, Canada.}

\author{Al\'{a}n Aspuru-Guzik}
\email[E-mail:]{aspuru@utoronto.ca}
\affiliation{Chemical Physics Theory Group, Department of Chemistry, University of Toronto, Canada.}
\affiliation{Department of Computer Science, University of Toronto, Canada.}
\affiliation{Department of Chemical Engineering and Applied Chemistry,  University of Toronto, Canada.}
\affiliation{Department of Materials Science and Engineering, University of Toronto, Canada.}
\affiliation{Vector Institute for Artificial Intelligence, Toronto, Canada.}
\affiliation{Canadian  Institute  for  Advanced  Research  (CIFAR)  Lebovic  Fellow,  Toronto,  Canada.}
\date{\today}

\begin{abstract}
    As we continue to find applications where the currently available noisy devices exhibit an advantage over their classical counterparts, the efficient use of quantum resources is highly desirable.
    The notion of quantum autoencoders was proposed as a way for the compression of quantum information to reduce resource requirements.
    Here, we present a strategy to design quantum autoencoders using evolutionary algorithms for transforming quantum information into lower-dimensional representations.
    We successfully demonstrate the initial applications of the algorithm for compressing different families of quantum states. 
    In particular, we point out that using a restricted gate set in the algorithm allows for efficient simulation of the generated circuits.
    This approach opens the possibility of using classical logic to find low representations of quantum data, using fewer computational resources.
\end{abstract}

\maketitle

\section{Introduction}
The availability of the noisy intermediate-scale quantum (NISQ)~\cite{preskill2018quantum} devices has spurred significant interest in finding different applications~\cite{bharti2022noisy, cerezo2020variational, cao2019quantumreview, mcardle2020quantumreview, Anand2021Quantum} for these devices.
Many proposals have been put forward in this regard.
However, the limited coherence times and connectivity of the available devices~\cite{arute2019quantum, ibmq, pino2020demonstration, jurcevic2020demonstration, rudolph2017optimistic} restrict the implementation of most proposed algorithms.
As such, different ways~\cite{peruzzo2014variational, farhi2014quantum, McClean2016theoryofvqe} to use the devices efficiently have been explored in the past.
One such direction is using an autoencoder ~\cite{goodfellow2016deep} to create a low-dimensional representation of the associated quantum states, allowing for noise-robust simulations~\cite{sharma2020noise, anand2021noise}.

An autoencoder is a popular tool for dimensionality reduction in classical data processing, as well as generative modeling.
However, using the classical model for compressing quantum data can lead to prohibitive resource requirements.
So, a quantum autoencoder (QAE) was proposed in Ref.~\cite{romero2017quantum} and Ref.~\cite{wan2017quantum} as a way to reduce this cost by using a quantum model for the task.
It uses a hybrid quantum-classical optimization loop to compress quantum states and has been used for a variety of tasks including, quantum simulation~\cite{romero2017quantum} and denoising quantum data~\cite{bondarenko2020quantum}, among others~\cite{pivoluska2022implementation, huang2020realization, hebenstreit2017compressed, pepper2019experimental, cao2021noise, locher2022quantum}.
All these studies rely on the ability of the chosen parameterized unitary circuit to compress the input quantum states, which becomes harder with the problem size and could not be generalized for the proposed application so far.
In a recent work~\cite{bravo2021quantum}, the authors modify the original QAE by including a feature vector characterizing the data and show that one can achieve compression with higher fidelity using their method.
While this is an important step, there is still a need for a general tool that can be used to design models for the compression of quantum data.

Recent studies~\cite{rattew2019domain, bilkis2021semi, chivilikhin2020mog, du2020quantum, zhang2020differentiable} have investigated the use of different algorithms to automatically search optimal circuit structures for different applications.
A tool that has been used to design circuit architectures from scratch successfully is evolutionary algorithms~\cite{rechenberg1978evolutionsstrategien, schwefel1977numerische}.
The idea of using evolutionary algorithms for quantum computing is not new~\cite{las2016genetic}, and more recently there has been considerable interest in the topic.
They have been used to generate ans\"atze for variational quantum eigensolver (VQE)~\cite{rattew2019domain},  design feature maps for classification~\cite{altares2021automatic}, and optimize variational algorithms~\cite{anand2021natural}, among others~\cite{Lamata_2018, zhao2020natural, bilkis2021semi, chivilikhin2020mog, du2020quantum, zhang2020differentiable}.

\begin{figure*}[tbp!]
  \centering
   \includegraphics[width=1.65\columnwidth]{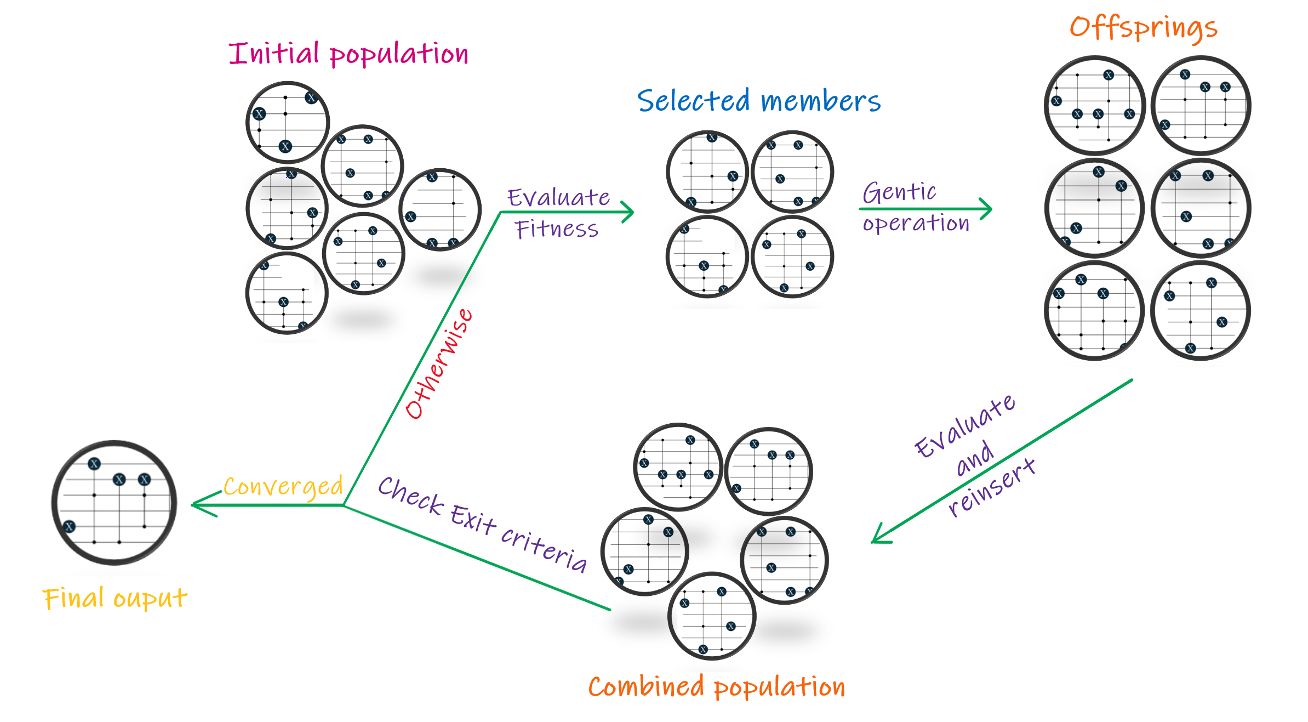}
   \caption{A figure showing the basic steps used in the evolutionary algorithm.}
   \label{fig:GA_outline}
\end{figure*}

In this article, we propose a general tool to design ans\"atze for compressing quantum data.
We chose evolutionary algorithms from the different classical algorithms to search for optimal circuit structures.
The algorithm uses a restricted set of gates (NOT, CNOT and Toffoli) that can be simulated efficiently on a classical computer to sequentially compress quantum data.
This implies that one can use our strategy to search for circuit structures that transform quantum data to a lower-dimensional representation efficiently.
We show the initial applications of the algorithm by finding low-dimensional representations of different families of quantum states.
Our work provides a simple and efficient approach for designing quantum models for compressing quantum data.
Furthermore, this will allow for encoding quantum states into low-dimensional representation using a classically tractable method, for simulating larger problems using the already existing algorithms~\cite{aspuru2005simulated, peruzzo2014variational, biamonte2017quantum}. 

The rest of the paper is organized as follows: we provide some background information about relevant topics in section~\ref{section:prelims}, the details of our compression algorithm in section~\ref{sec:method}, the result from numerical simulation in section~\ref{sec:simulations}, and finally present some concluding remarks in section~\ref{sec:conclusion}. 

\section{Preliminaries}\label{section:prelims}

\subsection{Evolutionary Algorithm}
Evolutionary algorithms~\cite{eiben2015evolutionary, whitley2001overview, zhou2011multiobjective} are heuristic search methods based on Darwin's theory of evolution \cite{darwin2004origin} which provide a robust and powerful way to find global solutions to complex optimization problems. 
The algorithm mainly contains four overall steps: initialization, selection, genetic mutations, and termination.
An initial set of solution candidates, often referred to as the population is chosen randomly or based on known solutions.
The candidates are generally represented using strings (programs) to facilitate mutation operations to be carried out on them.
During the selection step, the algorithm requires a quantitative measure of how close the current result is to the overall goal, usually referred to as the fitness function.
The candidates that show the most promise (determined by the value of the fitness function) are carried forward or used to create children for the next generation using different genetic mutations.
The selection and mutation steps are repeated until a solution is found.
A schematic of the main steps of an EA is shown in Figure~\ref{fig:GA_outline}.

\subsection{Compression of Quantum Data}
A quantum autoencoder (QAE) is a variational algorithm proposed in Ref.~\cite{romero2017quantum} for the compression of quantum data on a quantum computer. 
The goal is to find a unitary operation $U$ that compresses quantum data (set of quantum states $\{ \ket{\psi_i} \}$) to a latent representation, $U: \mathcal{H}^n \rightarrow \mathcal{H}^k \otimes \mathcal{H}^{n-k}$. 
The quantum states are transformed as: $\ket{\psi_i} \rightarrow \ket{\phi_i}\otimes\ket{x}$, where $\ket{\phi_i}$ is the low-dimensional representation on $k$-qubits and $\ket{x}$ is the trash state independent of the state $\ket{\psi_i}$, set to $\ket{x} = \ket{0}^{\otimes (n-k)}$ in our case.
The QAE consists of two networks, an encoder $U$ and a decoder $U^{\dagger}$, which maps quantum data to a latent space and back to the original representation. 
The training over test states $\{ \ket{\psi}_i \}$ is carried out minimising the following cost function:
\begin{align}
    \mathcal{L} = - \sum_i p_i \langle I^k \otimes \ket{0}\bra{0}^{n-k} \rangle_{U \ket{\psi_i}},
\end{align}
where, $p_i$ is the weight associated with the compression of every state,  and $U$ is the unitary that transforms the state $\ket{\psi_i}$ to a compressed representation as $U\ket{\psi_i} = \ket{\phi_i} \otimes \ket{0}^{\otimes (n-k)}$.
Current implementations of the QAE employ fixed parameterized ans\"atze for achieving optimal compression.
The success of the compression depends on the choice of the unitary $U$, as the circuit should be able to disentangle the two-qubit sets for all training states, which is hard in general for a polynomial depth circuit. 

An alternative approach to the above proposal is the use of adaptive protocols to construct ans\"atze, as proposed in Refs.~\cite{grimsley2019adaptive, tang2019qubit, ryabinkin2018qubit, ryabinkin2020iterative, lang2020iterative} for estimating eigenvalues.
We further build on the adaptive approach and propose a strategy for finding unitaries that can be simulated classically, to disentangle qubits sequentially to achieve optimal compression.
The details of our strategy is presented in section~\ref{sec:method}.

\subsection{Clifford Gates and Classical (Quantum) Logic}\label{sec:classical_quantum_logic}
The Clifford gates acting on $n$ qubits are elements of the Clifford group $C_n$ that normalize the corresponding Pauli group $P_n$, i.e., an unitary $U$ is a Clifford gate if for all $P \in P_n$, $P^{'} = U P U^\dag$ is also an element of the Pauli group, $P^{'} \in P_n$. 
Some common Clifford gates are the Pauli gates ($I$, $X$, $Y$, $Z$), phase-gate ($S$), Hadamard gate ($H$), Controlled-$Z$ gate ($CZ$), CNOT ($CX$) gate and SWAP gate, among others. 
The Clifford gates do not form a universal set of quantum gates. 
However, when augmented by the $T$ gate, they are universal for quantum computation.

Another common gate that is not a member of the Clifford group is the Toffoli (CCX) gate. 
It (together with initialized ancilla bits) is a reversible logic gate that is universal for classical logic, i.e., any classical logical operation can be decomposed into only Toffoli gates.
It alone, however, is not a universal gate for quantum computation.

It is known from the Gottesman-Knill theorem~\cite{gottesman1998heisenberg}, that any circuit that consists of only Clifford gates can be simulated efficiently using a classical computer.
Also, the action of just a Toffoli gate on arbitrary states can be efficiently simulated, given efficient access to the initial state.
This is because the action of a Toffoli cannot increase or decrease the number of basis states in a given quantum state.
Thus, the action of the gate is just updating the basis states present in the given state by looking up a truth table, which can be done efficiently if we are given efficient access to the state.

The action of a circuit consisting of a subset of the gates mentioned above can be simulated efficiently, and we refer to this as classical quantum logic.
One such subset is the set, $X$, $CX$, and $CCX$ gates, which are not universal for quantum computation. 
We will use this subset for the design of quantum circuits for the compression of quantum states.

\subsection{Circuit representation}
\begin{figure}[htbp]
    \subfloat[An example circuit built from the restricted gate set used in the algorithm.]{\includegraphics[width=0.95\linewidth]{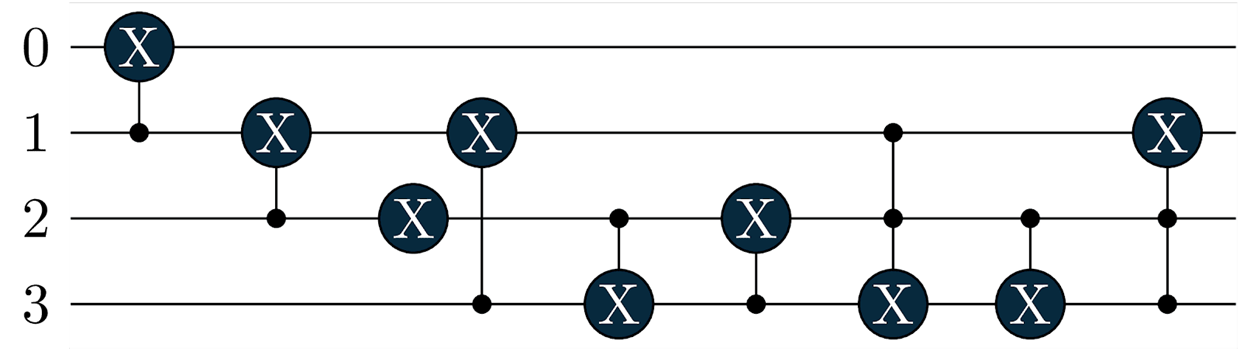}}\\
    \subfloat[The string representation of the circuit shown above.]{\includegraphics[width=0.95\linewidth]{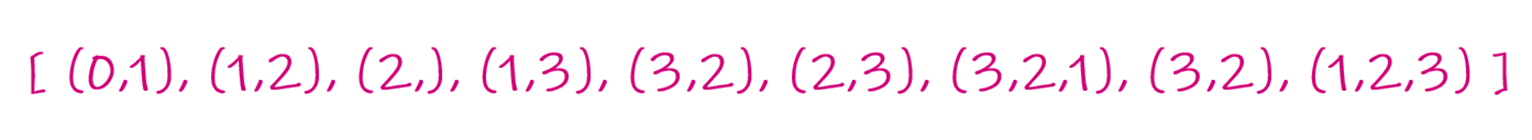}}
    \caption{A figure showing the string representation of a circuit use in the algorithm.}
    \label{fig:str_rep}
\end{figure}

An important aspect of the proposed algorithm is the efficient storage of the circuits used. 
We use a very simple strategy of representing the circuits as strings for efficient manipulation during the evolution.
The gates used for the design of the circuits are represented as a tuple of qubits it acts on, and thus the length of the tuple for the gates $X$, $CX$ and $CCX$ are 1, 2 and 3, respectively. 
The target qubit is the first element of the tuple followed by the control qubit(s).
The circuits are then represented by concatenating the string representation of the gates by sorting them by the qubits on the which they act and moments.
The gate encoding scheme is detailed in Table~\ref{tab:str_rep}, and an example string representation of a circuit is shown in Figure~\ref{fig:str_rep}.

\begin{table}[htbp]
    \centering
    \begin{tabular}{c|c}
         \hline
         Gates & String representation \\
         \hline
         \hline
         NOT ($X$) &  ($q_t$) \\
         CNOT ($CX$) &  ($q_t, q_c$)\\
         TOFFOLI ($CCX$) &  ($q_t, q_{c1}, q_{c2}$) 
    \end{tabular}
    \caption{A table summarizing the string representation of the gates used in the algorithm. The symbols $q_t$ represent the target qubit on which the gate is acting, while the symbols $q_c$, $q_{c1}$, and $q_{c2}$ represent the control qubits of the gates.}
    \label{tab:str_rep}
\end{table}

\section{Method}\label{sec:method}
We propose a method that allows for adaptive construction of an ansatz to transform different families of quantum states into a compressed representation.
The procedure uses an Evolutionary algorithm for generating circuits that disentangles a target qubit from the rest of the qubits in the input state. 
An outline of the evolutionary algorithm used is shown in Algorithm~\ref{al:ea}. 

\medskip

\begin{algorithm}[htbp]\label{al:ea}
\SetAlgoLined
	\textbf{input}: target qubit to disentangle, input state \\
	\textbf{initialize}: Generate a set of unitaries \{$U_1$, $U_2$, .. , $U_n$\}\\
    \While{stopping criterion is not met}
     {
      1. calculate the fitness ($f_1, f_2, \cdots , f_n $) of the unitaries in the set and rank them as per their fitness values. \\
      2. select a set of unitaries ($U_1, U_2, \cdots , U_m$) to perform genetic operations on. \\
      3. calculate the fitness ($f_1^{'}, f_2^{'}, \cdots , f_y^{'}$) of the new unitaries ($U_1^{'}, U_2^{'}, \cdots , U_y^{'}$). \\
      4. select new unitaries ($U_1^{'}, U_2^{'}, \cdots , U_x^{'}$) to add to the set of fit unitaries ($U_1, U_2, \cdots , U_m$). \\
      5. check the stopping criteria.
      }
 \caption{An outline of the Evolutionary Algorithm for disentangling a qubit.}
\end{algorithm}
\medskip 

The fitness function used in the algorithm is the overlap of the state of the target qubit with zero state. 
The expression of the fitness for a unitary $U_i$ for a set of input states $\{ \psi_j \}$ is written as:
\begin{equation}
    f_i = \sum_j \langle 0 | \Psi^{(t)}_j \rangle = \sum_j \langle 0 | (U_i \psi_j)^{(t)} \rangle,
\end{equation}
where, $\Psi_j^{(t)}$ is the state of the target qubit after the action of the unitary $U_i$ on the input state $\psi_j$, i.e., $\Psi_j = U_i \psi_j$.
The stopping criteria is finding an unitary that maximizes the fitness function, i.e., $f_i = 1$.

The different genetic operations that are allowed during the evolution are listed below:
\begin{enumerate}
    \item Addition of gates: A random gate from the gate set is added to a random position in the circuit.
    \item Removal of gates: Some gates is randomly picked and removed from the circuit.
    \item Permutation of gates: The gates in the circuit are arranged according to a random permutation of the previous positions.
    \item Repeat gates: Some gates are randomly picked and added to the circuit.
    \item Replacing gates: Some gates are randomly picked and replaced with another randomly picked gates.
\end{enumerate}

\begin{figure}[htbp!]
    {\includegraphics[width=0.99\linewidth]{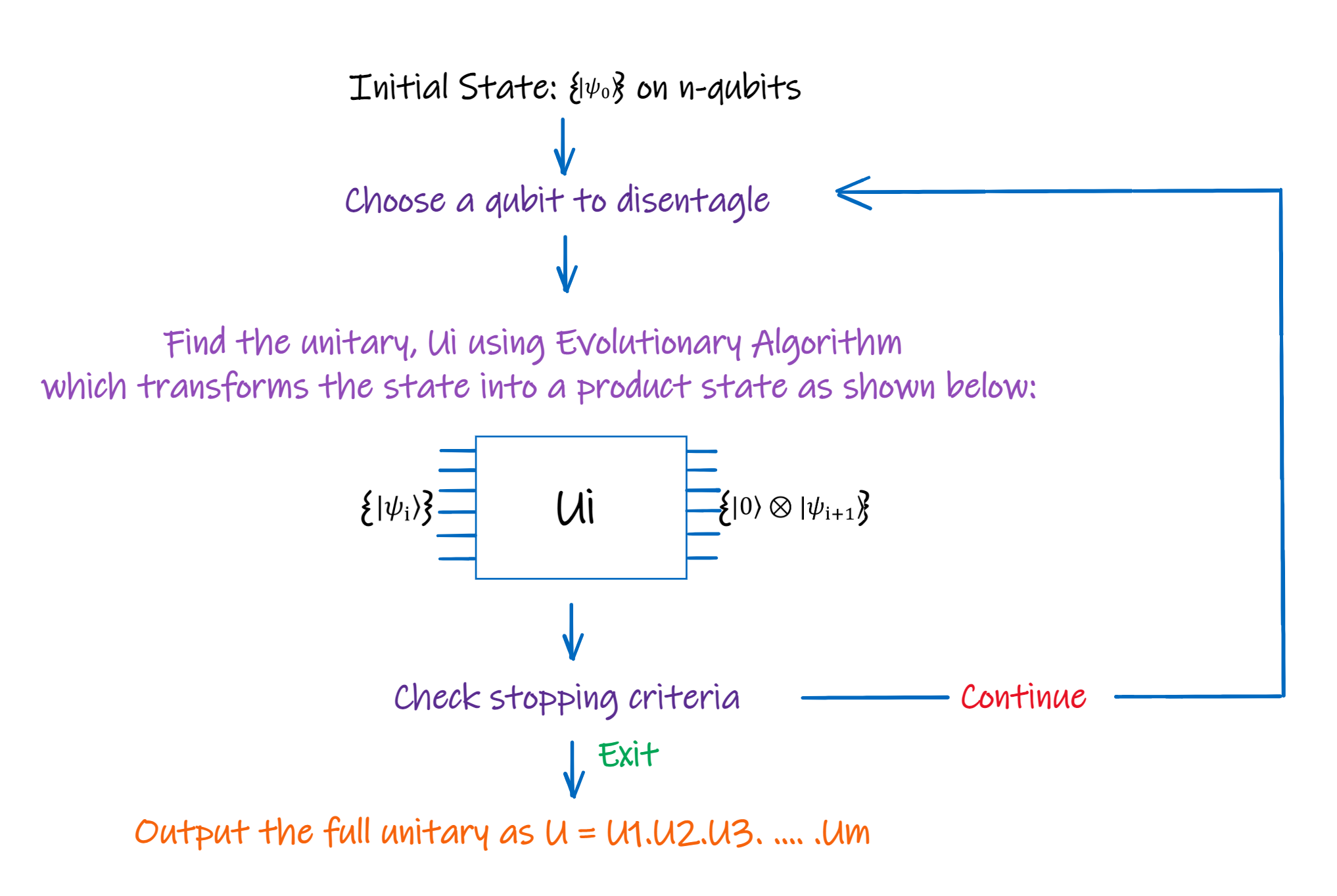}}
    \caption{An overview of the full algorithm for compression of a set of quantum states.}
    \label{fig:overall_algo}
\end{figure}

\begin{figure}[htbp!]
    {\includegraphics[width=0.99\linewidth]{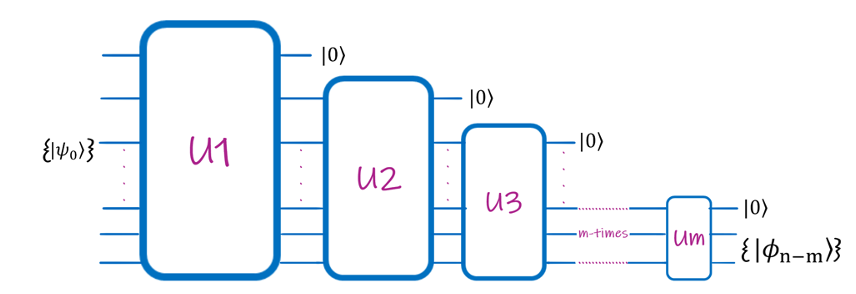}}
    \caption{A picture representing the full circuit constructed using the full algorithm.}
    \label{fig:full_circ}
\end{figure}

The above process for disentangling a qubit is repeated until the initial state has been transformed into a compressed representation. 
An illustration of the overall process is presented in Figure~\ref{fig:overall_algo}.
At the end of the overall circuit is the constructed by adding the individual circuits from the different repetitions as shown in Figure~\ref{fig:full_circ}.

The EA only uses the restricted set of gates, $X$, $CX$ and $CCX$, which allows for efficient simulation of all the circuits designed in the process as discussed in section~\ref{sec:classical_quantum_logic}.
Thus, the algorithm allows for a way of finding a unitary circuit that can transform a set of quantum states into a compressed representation.

\begin{figure*}[htbp!]
    {\includegraphics[width=0.4\linewidth]{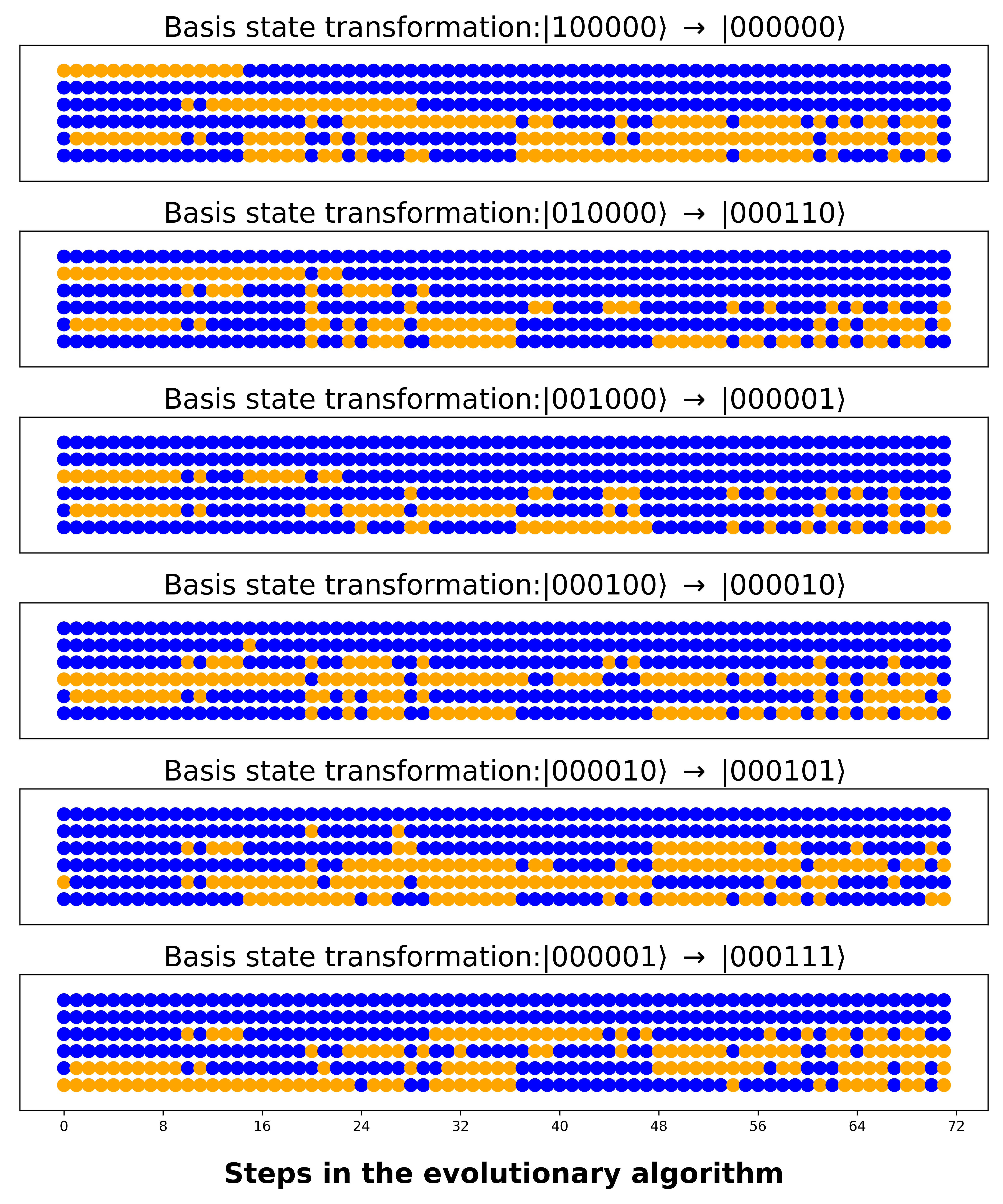}}
    \hspace{0.065\linewidth}
    {\includegraphics[width=0.4\linewidth]{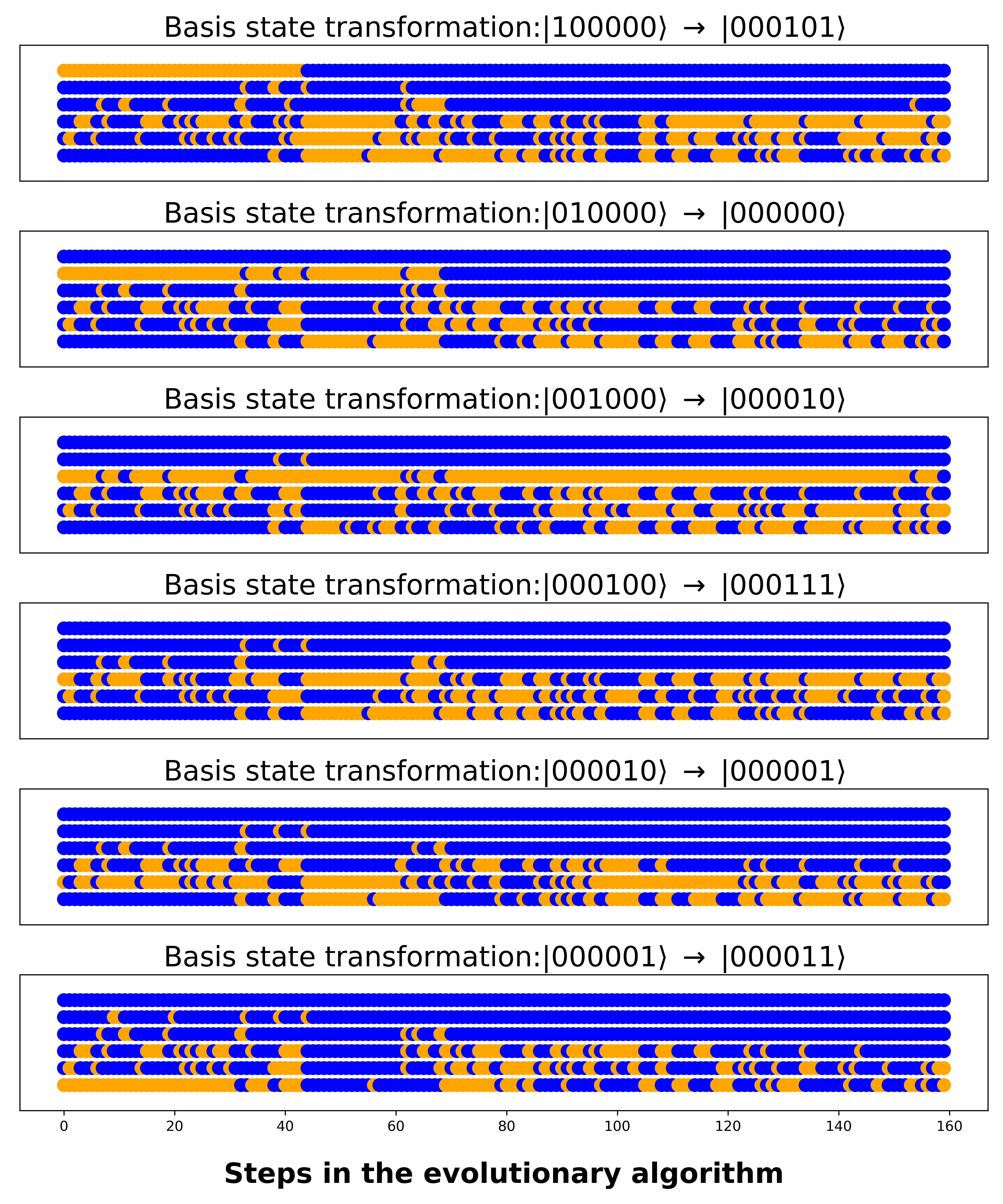}}
    \caption{An illustration of the evolution of the basis states in two different runs of the 6-qubit unary to binary compression. The orange and blue colored dots represent the state `1' and `0', respectively. The states at each step corresponds to the transformation with the fittest unitary. The final circuit of the left evolution contains 6-X, 18-CX, and 17-CCX gates, whereas that of the right contains 8-X, 12-CX, and 14-CCX gates. At the end of both the runs, the first three qubits are compressed, i.e., in a product state $\ket{000}$. }
    \label{fig:6q-unary-binary}
\end{figure*}

\section{Numerical Experiments}\label{sec:simulations}
We numerically test the validity of our method on different families of states of varying sizes.
We use some of the very common families of states that are of interest to the community, such as the unary states (states supported only by computational basis states with a single qubit in state $\ket{1}$), GHZ states, 2 and 3 particle states (states supported by computational basis functions with only 2 or 3 qubits in state $\ket{1}$), prime states, and random states.

To motivate the use of the restricted gate set we begin by using a random search algorithm that design circuits for transforming unary states into their binary representation, as shown below:
\begin{align}\label{eq:un_bin}
    \ket{\Psi_i} &\rightarrow  \ket{\Phi_j} \otimes \ket{0}^{\otimes N - log_2 N} \nonumber \\
    &\equiv \ket{\phi_1 \phi_2 .. \phi_{log_2 N}} \otimes \ket{0}^{\otimes N - log_2 N} ,
\end{align}
where, $\ket{\Psi_i}$ is an arbitrary computational basis state in a unary state and $\ket{\Phi_j}$ is the low-dimensional representation on $(log_2 N)$-qubits, with $\phi \in \{ 0, 1\}$.

The random search algorithm is based on some prior knowledge of gate patterns that work for smaller problem instances.
This includes searching for specific operations on some qubit combinations first, starting from empirically known patterns, like a cascade of CNOT gates.
This works well for compression of unary states but fails for other states such as the 2 or 3 particle states.

We next used the algorithm presented in section~\ref{sec:method} to search for unitaries that can carry the same compression of the unary states. 
The motivation behind this is to have a systematic exploration of circuit structures instead of a random search. 
We tested states of varying size, 4-8 qubits, and were able to find unitaries that transform these unary states to their binary representation.
We ran the algorithm multiple times (given its heuristic nature) to look for different circuit structures that perform the same operation.
The states from two different runs (best and worst in terms of time steps) of the 6-qubit unary compression at different steps of the algorithm are shown in Fig.~\ref{fig:6q-unary-binary}.
As can be seen from the Figure, the algorithm finds different transformations to achieve the same goal.
The shortest circuit that the algorithm finds for the 6-qubit unary compression however contains 6-X, 8-CX, and 12-CCX gates, which takes $\sim 90$ steps.
Thus, one can further improve the training to get shorter circuits by introducing a term in the fitness function penalizing the longer circuits.

We also tested our algorithm on a trivial problem of finding unitaries to transform 4-8 qubit GHZ-states~\cite{greenberger1989going} to a single qubit state, as:
\begin{equation}
    \ket{0}^{\otimes N} + \ket{1}^{\otimes N} \rightarrow \ket{0} + \ket{1}
\end{equation}
The algorithm was able to construct the unitaries very easily as one would expect. 
To further test our algorithm we formulated a harder compression problem where the input state (a superposition of N-random N-qubit basis states) has to be transformed into the compact binary representation, as:
\begin{equation}
    \ket{\Psi_i} \rightarrow \ket{\Phi_j}
\end{equation}
where, $\ket{\Psi_i}$ is sampled randomly (without replacement) from the possible $2^N$ possible basis states, and $\ket{\Phi_j}$ is a low dimensional representation as defined above (Eq.~\ref{eq:un_bin}).
For the 
This can be of importance as most states that we encounter in reality are not a superposition of all the possible basis states but a subset of them.
Using the algorithm, we again can find unitaries that can carry the above transformation for different 4-8 qubit states.

We now move to test our algorithm to transform physically motivated states, such as the prime-states~\cite{latorre2013quantum} which can be used to study interesting theorems in number theory.
A n-qubit prime state is formed by superposition of prime numbers less than $2^n$, as shown below:
\begin{equation}
    \ket{\Psi_p} = \frac{1}{N} \sum_{j \in Primes < 2^n} \ket{j}
\end{equation}
where, N is the normalization constant.
We were able to find unitary circuits that transform prime states on 4-6 qubits to a compressed representation.

Furthermore, we test our method on different $m$-particle $n$-qubit states (m-qubits in $\ket{1}$ state and remaining qubits in $\ket{0}$ state), which represent general states of interest in different quantum chemistry applications.
First, we transform all possible 2-particle states on 4-6 qubits to their compressed representation.
An example transformation of the 5-qubit 2-particle basis states is shown below:
\begin{align}
    \ket{11000} \rightarrow \ket{00111} \nonumber \\ 
    \ket{10100} \rightarrow \ket{00011} \nonumber \\
    \ket{10010} \rightarrow \ket{00001} \nonumber \\
    \ket{10001} \rightarrow \ket{01100} \nonumber \\ 
    \ket{01100} \rightarrow \ket{01011}  \\
    \ket{01010} \rightarrow \ket{01001} \nonumber \\
    \ket{01001} \rightarrow \ket{00110} \nonumber \\
    \ket{00110} \rightarrow \ket{00101} \nonumber \\
    \ket{00101} \rightarrow \ket{00010} \nonumber \\
    \ket{00011} \rightarrow \ket{00000} \nonumber 
\end{align}

\begin{figure}[htbp]
    {\includegraphics[width=0.99\linewidth]{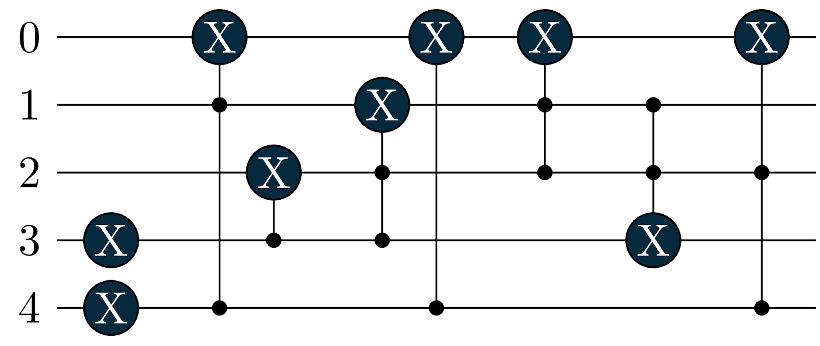}}
    \caption{A picture representing the circuit that transform the 5-qubit 2-particle states to a compressed representation.}
    \label{fig:5q_2p_circuit}
\end{figure}

The circuit designed by the algorithm to carry the transformation in (7) is shown in ~\ref{fig:5q_2p_circuit}. 
As can be seen in the transformation, we can use the above unitary to represent any 5-qubit 2-particle state using only 4 qubits, thus saving a qubit.

\begin{figure*}[htbp!]
  \centering
   \includegraphics[width=1.5\columnwidth]{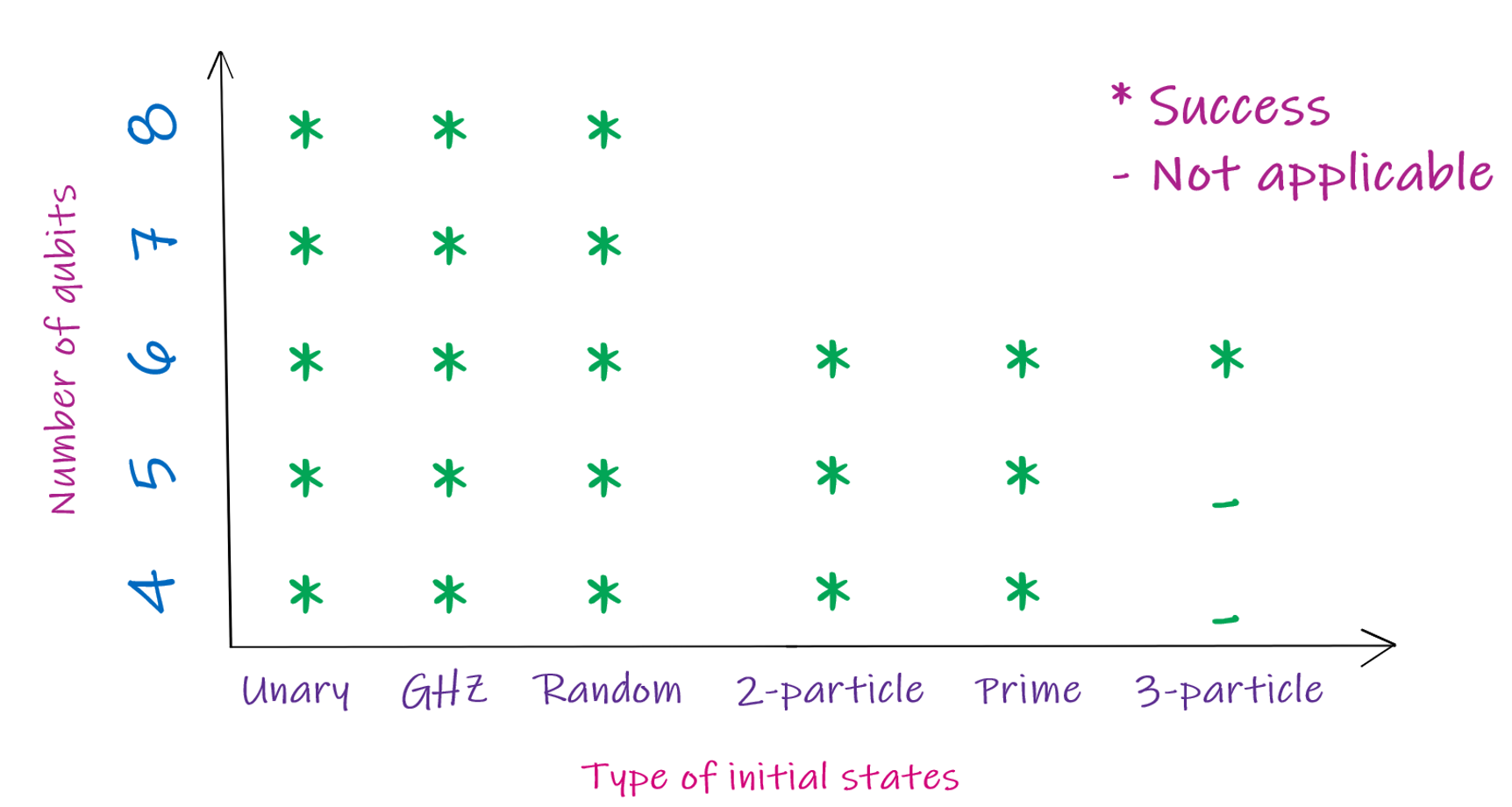}
   \caption{A figure showing the summary of results from the compression algorithm. The case marked with `-' are not applicable as they translate to a simpler compression problem, e.g., 5-qubit 3-particle state is equivalent to compressing 5-qubit 2-particle state. }
   \label{fig:GA_results}
\end{figure*}

Finally, we also try to find a circuit that transforms the 6-qubit 3-particle state to the corresponding compressed representation.
We again can find a unitary that transforms them to a 5-qubit state.
All the results from the numerical experiments are presented in Figure~\ref{fig:GA_results}, where we summarize the success of the algorithm in finding a unitary that can transform different states into a compressed representation. 
All the data corresponding to the circuits that carry out the different transformations can be found in this github repository.~\cite{Anand2022}

\section{Conclusion}\label{sec:conclusion}
In this article we have presented a method for finding unitaries that can transform different families of quantum states into a compact representation.
The algorithm carries a heuristic search of circuit structures that can sequentially disentangle qubits from a quantum state.
It uses a restricted gate set which is universal for classical logic operations, but not for quantum computation.
This allows for efficient simulation of the unitaries constructed from the set as the action of the unitaries on basis states can be calculated using a simple lookup table.

We then motivated the use of such a restricted gate set using a random structure search for finding unitaries that transform the unary state to its binary representation.
Though the random search works well in finding unitaries for the above task, it does not work well beyond this.
So, we move to a more structured approach and use an evolutionary algorithm to further find unitaries for transforming other families of quantum states.

We numerically present evidence of the success of the algorithm for finding unitaries that transform different states, such as unary states, GHZ states, random states, prime states, and n-particle states.
This study presents a simple procedure that can be used to find unitaries that transform a quantum state into a more compact representation and can be implemented efficiently on classical computers.
This can be of utmost interest to the scientific community as we push toward finding useful applications that can be implemented on the available quantum devices.

Finally, this work also opens the pathway to exploring the use of only classical logic operations to carry out compression of quantum data, as well as the use of restricted gate sets to construct circuits for correlating different subsystems.
We leave this to future study, as it requires additional work.

All the code and data supporting the results in the article are available on github~\cite{Anand2022}. 

\section*{Acknowledgements}
The authors thank Riley J. Hickman and Akshat K. Nigam, Alba Cervera-Lierta and Matthias Degroote for helpful discussions about the project and Philipp Schleich for providing valuable comments regarding the manuscript.
This work was supported by the U.S. Department of Energy under Award No. DE-SC0019374.
A.A.-G. acknowledges the generous support from Google, Inc.  in the form of a Google Focused Award.
A.A.-G. also acknowledges support from the Canada Industrial Research Chairs  Program and the Canada 150 Research Chairs Program. Computations were performed on the niagara supercomputer at the SciNet HPC Consortium.~\cite{niagara1, niagara2} SciNet is funded by: the Canada Foundation for Innovation; the Government of Ontario; Ontario Research Fund - Research Excellence; and the University of Toronto.
We thank the generous support of Anders G. Fr\o{}seth.

\bibliography{main.bib}

\end{document}